# Interfacial thermal conductance across metal-insulator/semiconductor interfaces due to surface states

Tingyu Lu,[1] Jun Zhou,[1*] Tsuneyoshi Nakayama,[1,2] Ronggui Yang,[3,4] and Baowen Li[3]

[1]Center for Phononics and Thermal Energy Science, School of Physics Science and Engineering, Tongji University, Shanghai 200092, People's Republic of China

[2]Hokkaido University, Sapporo 060-0826, Japan

[3]Department of Mechanical Engineering, University of Colorado, Boulder 80309, USA

[4]Materials Science and Engineering Program, University of Colorado, Boulder 80309, USA

**Abstract:** We point out that the effective channel for the interfacial thermal conductance, the inverse of Kapitza resistance, of metal-insulator/semiconductor interfaces is governed by the electron-phonon interaction mediated by the surface states allowed in a thin region near the interface. Our detailed calculations demonstrate that the interfacial thermal conductance across Pb/Pt/Al/Au-diamond interfaces are only slightly different among these metals, and reproduce well the experimental results of the interfacial thermal conductance across metal-diamond interfaces observed by Stoner *et al.* [Phys. Rev. Lett. **68**, 1563 (1992)] and most recently by Hohensee *et al*. [Nature Commun. **6**, 6578 (2015)].

[*]To whom correspondence should be addressed. Email: zhoujunzhou@tongji.edu.cn

**I. INTRODUCTION**

Kapitza resistance, which occurs when heat flows across two different materials was first observed by the measurement of the temperature jump at solid-liquid helium interfaces.[1,2] Interfacial thermal conductance (ITC), the inverse of Kapitza resistance, between solid-solid interfaces has been the most intensive research topic in thermal transport community over the past two decades for thermal management of nanoelectronics and for engineering nanostructured materials for thermoelectrics and for superior thermal insulators.[3-7] Overheating caused by heat accumulation due to the presence of Kapitza resistance is a major obstacle to improving the performance and the reliability of nanoelectronic and optoelectronic devices. It is of great importance to understand the underlying mechanisms of Kapitza resistance or ITC across solid-solid interfaces.

Comparing to nonmetal interfaces, the heat conduction mechanism across metal-insulator/semiconductor interfaces is much more complicated since it involves energy conversion and coupling among different energy carriers where electrons are the major heat carriers in metal and phonons are the major heat carriers in insulator/semiconductor. The applicability of various proposed mechanisms has been debated for decades.[5]

Swartz and Pohl[5] proposed diffusive phonon scattering at interfaces as an extension of the acoustic mismatch theory for solid-liquid interfaces proposed by Khalatonikov,[8] which takes into account the elastic scattering of phonons by the roughness of interfaces. However, the predicted ITC values based on this theory is quite different from the measured values at room temperature. Stoner *et al.*[9] measured ITC across a large series of metal (Pb/Ti/Al/Au)-insulator (diamond/sapphire/BaF$_2$) interfaces using picosecond laser-based pump-and-probe technique. They found that ITC across Pb-diamond interface was significantly larger than the radiation limit of phonon transmission, which considers the maximum possibility of phonon transmission. To overcome the underestimation of ITC, multi-phonon scattering due to anharmonicity was then proposed, which provides additional energy transfer channel across an interface. However, one would expect ITC to be reduced in the presence of pressure if anharmonicity is important, as the anharmonicity of diamond would be suppressed by pressure.[10] This prediction contradicts with the recent experimental exploration where pressure-dependent ITC across metal-diamond interface was measured by Hohensee et al.[11]

On the other hand, due to the existence of electrons in the metal side, the role of the electron-phonon (e-p) interaction between the electrons in metal and phonons in nonmetal has been considered, which was first introduced for metal-helium interfaces and then extended to metal-insulator/semiconductor interfaces. The theoretical framework of e-p interaction of metal-helium interface was proposed by Little[12] and Andreev[13] in 1960's and experimentally proved by Wagner et al.[14] through the observation of ITC dependence on the applied magnetic field.

Energy coupling between electrons and phonons in bulk materials was modeled by Kaganov[15] and Allen[16] several decades ago. For the e-p interaction of metal-insulator/semiconductor interfaces, there are several different models which show different mechanisms of interactions. Huberman and Overhauser[17] calculated the ITC across Pb-diamond interface by taking into account the coupling between the free electrons in Pb and joint vibrational modes near the interface in 1990's. In their work, they assumed that the phonon modes of an insulator can extend into the metal side with an attenuation rate. Sergeev[18] has proposed a model to incorporate the inelastic electron-boundary scattering in a similar way as the inelastic electron-impurity scattering. This model usually overestimates the ITC since all the atoms on the insulator/semiconductor side are treated as impurities. Mahan[19] has proposed a mechanism for the e-p interaction by introducing image charges which is likely inapplicable to non-polar materials such as diamond. Giri et al.[20] considered the electronic transmission probability, which is a fitting parameter, to calculate ITC. One important observation is that most of the above theories, except Mahan's, predict very strong dependence of ITC on electronic structure, which is in contradiction with experimental results.[9,11,21] For example, the calculated ITC across Pb-diamond interface is two orders-of-magnitude larger than that across Bi-diamond interface, whereas very similar measured values have been obtained.[21] Most recently, Hohensee et al.[11] have measured the ITC across interfaces between various metals (Pb/Pt/Al/Au) and diamond at high pressure up to 50 GPa. They found that ITC converge to similar values at high pressures among Pb/Pt/Al/Au.

The discrepancy between theoretical prediction and experimental observation of ITC across metal-insulator/semiconductor due to e-p interaction between the electrons in metal and phonons in insulator/semiconductor is still an outstanding question. Most theories assumed that free electrons from the metal side incident on the interface and are then reflected back while interacting with phonons at an interface.[17,18,20] This assumption could be wrong in realistic materials.

In this paper, we propose a new mechanism that determines e-p interaction across metal-insulator/semiconductor interfaces near room temperature, which is mediated by the surface states (SS).[22] The SS are localized electron states in an insulator/semiconductor induced by electrons from the metal side, whose wave-functions decay exponentially from the interface[23,24] with a decay length around several angstroms. This heat conduction channel due to the interaction of SS electrons with phonons in insulator/semiconductor side, which is noted as SS-phonon interaction in this paper, should be considered in parallel with phonon transmission-mediated thermal conductance.[25,26] When this channel dominates, we find that ITC across Pb/Pt/Al/Au-diamond interfaces vary only slightly among different metals with large difference in electronic structures, which agree well with the experimental results.[9,11,21] This theory also gives a sound explanation for the recently measured ITC under high pressures.[11]

This paper is organized as follows. We present the model to calculate ITC due to the SS-phonon interaction in Sec. II. This model is then used to calculate the ITC across Pb/Pt/Al/Au-diamond interfaces in comparison with experimental data from literatures in Sec. III. Finally, Sec. IV concludes this paper.

## II. MODEL

The SS exist in a very thin region near the interface that should play a crucial role in the energy exchange between the electrons in the metal and phonons in insulator/semiconductor. To simplify the modeling, we chose diamond as an example in this study. Our model can be easily extended to other materials when SS are important.

Figure 1(a) shows the band alignment of a metal-diamond interface where $E_F$ is the Fermi energy, $\Phi_M$ is the work function of metal, $E_g$ is the band gap with $E_0$ as its center, $\chi$ is the electron affinity, $\Phi_B$ is the Schottky barrier height, $E_{vac}$ is the vacuum energy level, $E_c$ and $E_v$ are the conduction and valence band edges, respectively. The band bending near the surface originates from the charge transfer between metal and diamond where the p-type diamond with boron doping is chosen as an example. $E_F$ is usually pinned at the surface to the SS.[27]

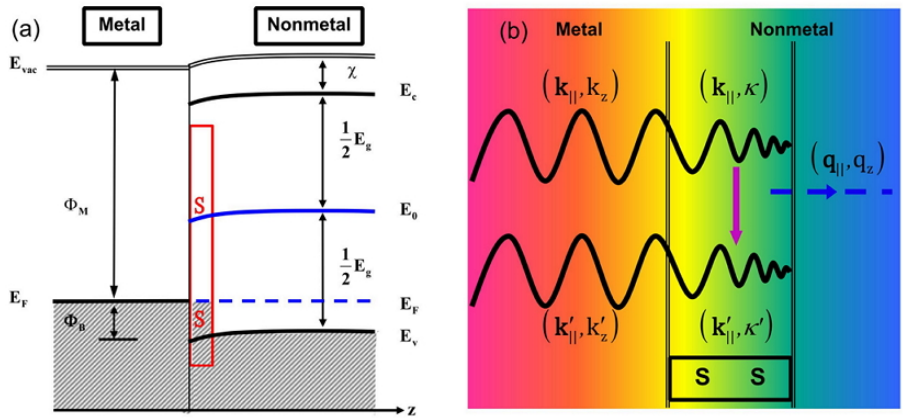

*FIG. 1. (color online) (a) Band alignment of a metal-diamond interface along the z-direction. The SS only exists in a thin interfacial region as shown. (b) Schematic of phonon emission due to SS-phonon interaction. $(q_{||}, q_z)$ denotes the wave vector of phonon, where $q_{||}$ is in x-y plane and $q_z$ is in z-direction. $(k_{||}, \kappa)$ and $(k'_{||}, \kappa')$ are the wave vectors of initial and final electron states of SS, respectively. $(k_{||}, k_z)$ and $(k'_{||}, k'_z)$ are the related wave vectors of electrons in metal.*

The temperature of the electrons in SS could be different from the temperature of electrons near the surface in metal due to the existence of interfacial electronic thermal resistance. In order to simplify the investigation, we neglect here the difference between these two temperatures since the electronic thermal resistance is believed to be smaller than the resistance due to SS-phonon interaction. The net heat flux from SS electrons to phonons in insulator/semiconductor due to SS-phonon interaction is given by $\Delta \dot{Q}_{NM}$, where $T_e$ and $T_p$ are the electron and phonon temperatures, respectively.[2,17] Then the ITC ($h_K$) due to SS-phonon interaction is defined by

$$h_K = \Delta \dot{Q}_{NM}/\Delta T; \quad \Delta T = T_e - T_p \ll T_e, T_p. \quad (1)$$

$\Delta \dot{Q}_{NM}$ defined in Eq. (1) can be calculated by[19,28]

$$\Delta \dot{Q}_{NM} = \frac{2\pi}{\hbar} \sum_{\lambda, \mathbf{k}_{||}, \kappa, \kappa', \mathbf{q}_{||}, q_z} \hbar \omega_{\lambda, q_{||}, q_z} |M_\lambda(q_{||}, q_z)|^2 |I(k'_{||}, k_{||}, \kappa', \kappa, q_z)|^2 W, \quad (2)$$

where $\hbar$ is the Planck constant, $\hbar \omega_{\lambda, q_{||}, q_z}$ is the phonon energy with polarization index $\lambda$ in insulator/semiconductor, $M_\lambda(q_{||}, q_z)$ is the scattering matrix elements and $I(k'_{||}, k_{||}, \kappa', \kappa, q_z)$ is the form factor. The transition probability $W$ is given by

$$W = \delta\left(\varepsilon_{\mathbf{k}_{||},\kappa} - \varepsilon_{\mathbf{k}_{||}+\mathbf{q}_{||},\kappa'} - \hbar\omega_{\lambda,q_{||},q_z}\right)\left[f_{\mathbf{k}_{||},k_z}\left(1 - f_{\mathbf{k}_{||}+\mathbf{q}_{||},\kappa'}\right)\left(n_{\lambda,q_{||},q_z} + 1\right) - f_{\mathbf{k}_{||}+\mathbf{q}_{||},\kappa'}\left(1 - f_{\mathbf{k}_{||},k_z}\right)n_{\lambda,q_{||},q_z}\right] -$$
$$\delta\left(\varepsilon_{\mathbf{k}_{||},k_z} - \varepsilon_{\mathbf{k}_{||}+\mathbf{q}_{||},\kappa'} + \hbar\omega_{\lambda,q_{||},q_z}\right)\left[f_{\mathbf{k}_{||},k_z}\left(1 - f_{\mathbf{k}_{||}+\mathbf{q}_{||},\kappa'}\right)n_{\lambda,q_{||},q_z} - f_{\mathbf{k}_{||}+\mathbf{q}_{||},\kappa'}\left(1 - f_{\mathbf{k}_{||},k_z}\right)\left(n_{\lambda,q_{||},q_z} + 1\right)\right]. \quad (3)$$

Here $f_{k_{||},\kappa}(T_e) = \left\{\exp\left[\left(\varepsilon_{k_{||},\kappa} - E_F\right)/(k_B T_e)\right] + 1\right\}^{-1}$ is the Fermi-Dirac distribution function where $\varepsilon_{k_{||},\kappa}$ is the electron energy and $k_B$ is the Boltzmann constant, $n_{\lambda,q_{||},q_z}(T_p) = \left\{\exp\left[\left(\hbar\omega_{\lambda,q_{||},q_z}\right)/(k_B T_p)\right] - 1\right\}^{-1}$ is the Bose-Einstein distribution function.

The second quantized form of SS-phonon interaction Hamiltonian is written as

$$H = \sum_{\lambda,\mathbf{k}_{||},\kappa,\kappa',\mathbf{q}_{||},q_z} M_\lambda(\mathbf{q}_{||},q_z) I(k'_{||},k_{||},\kappa',\kappa,q_z) c^\dagger_{\mathbf{k}_{||}+\mathbf{q}_{||},\kappa'} c_{\mathbf{k}_{||},\kappa} \left[a_{\lambda,\mathbf{q}_{||},q_z} + a^\dagger_{\lambda,-\mathbf{q}_{||},-q_z}\right], \quad (4)$$

where $c^\dagger$ ($a^\dagger$) and $c$ ($a$) are the creation and annihilation operators of electrons (phonons), respectively. In Fig. 1(b), we show the phonon emission process when electron is scattered from state $(\mathbf{k}_{||},\kappa)$ to state $(\mathbf{k}'_{||},\kappa')$. We consider the phonon modes in insulator/semiconductor side because SS mainly exist in a thin region in insulator/semiconductor.

Several methods have been proposed to calculate SS, such as electron wave function matching, Green's function matching,[24] and self-consistent pseudopotential method.[22] We use the simplest one-dimensional electron wave function matching method by considering a periodic potential $V_0 \cos(gz)$ ($V_0 < 0$) along z-direction when electrons are assumed to be free in x-y plane.[24] $g = 2\pi/(a/2)$ is the reciprocal lattice vector of diamond along (001) direction with lattice constant $b=a/2$. Then the electron wave function in the band gap of insulator/semiconductor is $\psi_{\mathbf{k}_{||},\kappa}(\mathbf{r}) = S^{-1/2} B e^{i\mathbf{k}_{||}\cdot\boldsymbol{\rho}} \varphi(\kappa,z)$ for $z \geq 0$, where $S$ is the area of interface and $B$ is a normalization factor. $\varphi(\kappa,z) = e^{-\kappa z}\cos(gz/2 + \phi/2)$ where $\phi$ is a phase shift. The positional coordinate is $\mathbf{r} = (\boldsymbol{\rho},z)$. For a given $\kappa$, there are two energy states: $\varepsilon_{k_{||},\kappa} \approx E_0 + \frac{\hbar^2 k_{||}^2}{2m} \pm |\xi|$, where $E_0$ is the center of gap and $|\xi| = \sqrt{V_0^2 - \frac{\hbar^4\kappa^2 g^2}{2m}}$. We focus our study on the localized SS electron which require $-|V_0| < \xi < |V_0|$. The electron wave function in metal is written as $\psi^M_{\mathbf{k}_{||},k_z}(\mathbf{r}) = S^{-1/2} A e^{i\mathbf{k}_{||}\cdot\boldsymbol{\rho}} \sin(k_z z + \eta)$ for $z \leq 0$ where $A$ is the normalization factor, and $\eta$ is a phase shift. For given $k_F$, we have $k_z = \sqrt{k_F^2 - k_\parallel^2}$ and $k_z' = \sqrt{k_F^2 - |\mathbf{k}_\parallel + \mathbf{q}_\parallel|^2}$.

The form factor for free phonon modes is then written as $I(k_\parallel', k_\parallel, \kappa', \kappa, q_z) = BB'^* \int_0^\infty dz\, \varphi^*(\kappa', z) e^{iq_z z} \varphi(\kappa, z)$. Here, only the electrons at the insulator/semiconductor side ($z > 0$) contribute to the ITC and the contribution from the metal side is negligible. The reason is that the e-p interaction for z<0 is merely the conventional e-p interaction in metal which does not contribute to ITC directly.[29] One can easily modify $e^{iq_z z}$ in phonon wave function to $\sin(q_z z)$ to consider the localized phonon modes.[19] The detailed expressions of the form factors for both cases are shown in Appendix.

## III. RESULTS AND DISCUSSIONS

We now turn to the calculation of ITC due to the interaction of SS electrons with phonons across metal-diamond interfaces. We choose $|V_0| = E_g/2 = 2.74$ eV for diamond.[30] We calculate the SS-phonon interaction due to both acoustic and optical modes by employing the deformation potentials. For longitudinal-acoustic (LA) modes, the squared scattering matrix elements is $|M_{LA}(q_\parallel, q_z)|^2 = \hbar D^2 q/(2V\rho_0 v_l)$[31] where $D$ is the deformation potential constant of acoustic phonons, $\rho_0$ is the mass density, $v_l$ is longitudinal sound velocity, and $V$ is volume. The transverse-acoustic phonon modes are considered similarly by using the transverse sound velocity $v_t$. The optical-phonon scattering with polar interaction is irrelevant since diamond is non-polar material. The squared scattering matrix elements of electron-longitudinal optical (OP) phonon scattering with deformation potential is $|M_{OP}(q_\parallel, q_z)|^2 = \hbar(D_1 K)^2/(2V\rho_0 \omega_{OP})$[30] where $D_1 K$ is the optical deformation potential constant, and $\omega_{OP}$ is the optical phonon frequency. The contributions from the two transverse optical phonon modes are considered similarly. These parameters of diamond are used: $a = 3.567 \text{Å}$, $v_l = 1.82 \times 10^4 \text{m/s}$, $v_t = 1.23 \times 10^4 \text{m/s}$, $\rho_0 = 3.515 \text{g/cm}^3$, and $D = 8.7 \text{eV}$.[32]

Table I. Fermi wave vectors, Schottky barrier heights, and corresponding $E_F - E_0$ used in calculations.

| Parameters | Pb | Pt | Al | Au |
| --- | --- | --- | --- | --- |
| $k_F$ (Å$^{-1}$) | 1.58[33] | 1.6[34] | 1.75[33] | 1.21[33] |
| $\Phi_B$(eV) | 2.03[35] | 1.56[35] | 2.0[35] | 1.71[35] |
| $E_F - E_0$(eV) | -0.71 | -1.18 | -0.74 | -1.03 |

Figure 2 show the calculated ITC across Pb/Pt/Al/Au-diamond interfaces as a function of $E_F - E_0$ when $T_p = 273\text{K}$. ITC considering both free and localized phonon modes are shown in Fig. 2(a) and Fig. 2(b), respectively. Two different optical deformation potentials are considered. One is $D_1K = 2.1 \times 10^9 \text{eV/cm}$ which comes from transport property measurements,[36] and the other is $D_1K = 3.09 \times 10^9 \text{eV/cm}$ which comes from Raman experiments.[37] The Fermi wave vectors of metals used in calculation are shown in Table I. We find that ITC for all four interfaces vary with $E_F - E_0$ because the SS and the form factor are energy-dependent which will be shown later.

For most metal-insulator/semiconductor interfaces, $E_F$ is pinned at surface to the SS in band gap.[23] Such pinning effect is characterized by the Schottky barrier height which enables us to approximately deduce the realistic $E_F - E_0$ by $\Phi_B - E_g/2$ as shown in Fig. 1(a). The values of $\Phi_B$ of four interfaces and the corresponding $E_F - E_0$ are listed in Table I. We mark these $E_F - E_0$ values with vertical lines in Figs. 2(a) and 2(b). With the obtained $E_F - E_0$, we find that ITC for Pb/Pt/Al/Au-diamond interfaces vary from 39.7MW/(m²K) to 44.4MW/(m²K) for free phonon modes and from 58.3MW/(m²K) to 65.9MW/(m²K) for localized phonon modes as shown in Table II. The ITC with localized phonon modes are about 50% larger than that with free phonon modes. For both cases, the differences among different metals are found to be only slightly because of the similar Fermi wave vectors and Fermi energies in gap. This finding explains well why the ITC approaches to similar values at high pressure.[11]

We further compare ITC due to SS-phonon interaction with the parallel channel of phonon transmission-induced ITC which is calculated by DMM and by the phonon radiation limit in Table II. The experimental data are also shown. For Pb/Au-diamond interfaces, our calculated ITC are much larger than that from DMM and are in good agreement with experimental data. This comparison indicates that SS-phonon interaction dominates the ITC of these two interfaces and the contribution from phonon transmission is negligible. For Pt-diamond interfaces, our calculated ITC is comparable to that from DMM and the phonon radiation limit and they must be considered in parallel. The summation of the ITC from the two channels varies from 76 MW/(m²K) to 96.2 MW/(m²K) which is 50%~65% of the experimental data. For Al-diamond interface, our calculated ITC is 30%~50% of that from DMM. The summation of the ITC from the two channels varies from 174.4 MW/(m²K) to 195.9 MW/(m²K) which is close to the measured data from Ref. 11 and the maximum value in Ref. 38. This finding implies that both SS-phonon interaction and phonon transmission should be considered in parallel[25,26] for Pt/Al-diamond interfaces.

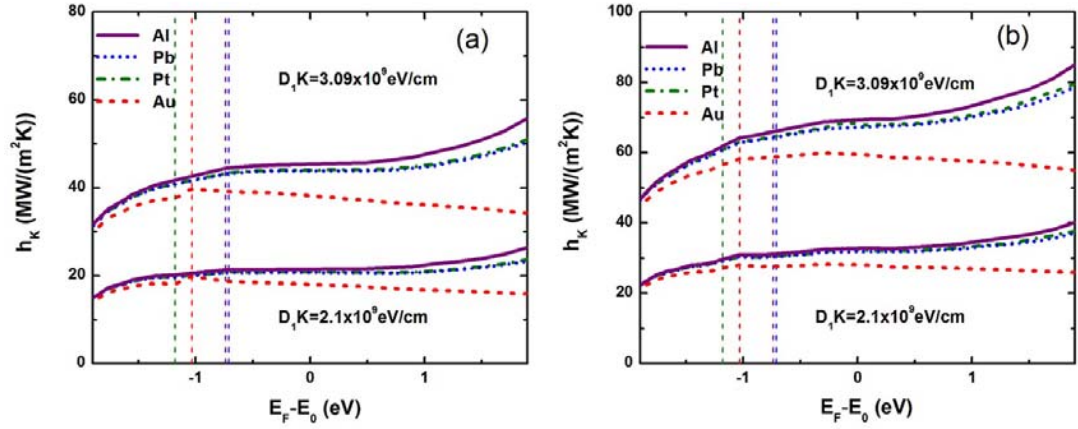

FIG. 2 (color online) Calculated ITC across Pb/Pt/Al/Au-diamond interfaces for (a) free phonon modes and (b) localized phonon modes as a function of $E_F - E_0$ for two different values of $D_1 K$ when $T_p$=293K. The realistic $E_F - E_0$ listed in Table I are marked with vertical lines and the corresponding ITC are shown in Table II.

Table II. Calculated ITC across Pb/Pt/Al/Au-diamond interfaces at room temperature for free and localized phonon modes when $D_1 K = 3.09 \times 10^9 \, \text{eV/cm}$. ITC from DMM and phonon radiation limit, and experimental measured values are listed for comparison. The unit is $[MW/(m^2 K)]$.

| ITC | Pb | Au | Pt | Al |
|---|---|---|---|---|
| SS-free phonon | 43.3 | 39.7 | 41 | 44.4 |
| SS-localized phonon | 64.5 | 58.3 | 61.2 | 65.9 |
| DMM | $2^9$ | $12^9$ | $35^{26}$ | $130^{26}$ |
| Radiation limits | $2.5^9$ | $24^9$ | $47^{11}$ | $207^{11}$ |
| Experimental data | $31^9, 60^{11}$ | $40^9, 62^{11}$ | $143^{11}$ | $46^9, 153^{11}, 23\text{-}180^{37}$ |

Figure 3 show the temperature-dependence of ITC across Pb/Pt/Al/Au-diamond interfaces when $D_1K = 3.09 \times 10^9 \text{eV/cm}$ and $E_F - E_0$ in Table I are used. Fig. 3(a) shows the dependence of ITC on $T_e$ when $T_p = 293K$. We find that $h_K$ increases with $T_e$ for all metals since more SS electrons participate the e-p interaction at higher temperature. This is quite different from the e-p coupling constant in bulk metals, which remains almost constant at room temperature.[19] Fig. 3(b) shows the dependence of $h_K$ on $T_p$. We find that $h_K$ increases with $T_p$ for all interfaces because the increased phonon population at higher $T_p$ leads to stronger e-p interaction and provides more energy transfer channels.

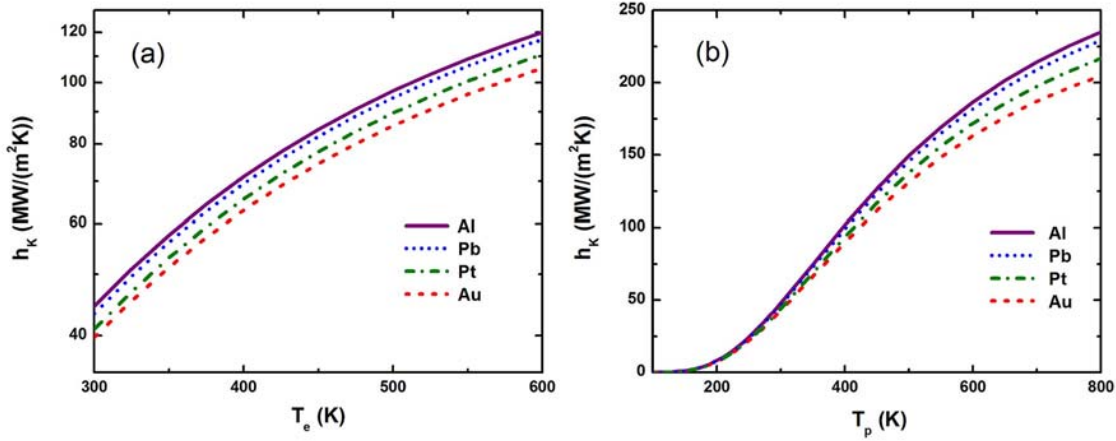

FIG. 3 (color online) Temperature dependence of ITC across Pb/Pt/Al/Au-diamond interfaces with $D_1K = 3.09 \times 10^9 \text{eV/cm}$. (a) When $T_p = 293K$, $h_K$ increases with $T_e$ because of more participating electron states at higher temperature. (b) $h_K$ dependence on $T_p$ for small $\Delta T$. The increased phonon population at higher $T_p$ leads to more energy transfer.

In order to better understand the dependence of ITC on Fermi energy and temperature, as shown in Figs. 2 and 3. We show the energy dependence of $\left|\psi_{\mathbf{k}_{||},\kappa}(\mathbf{r})\right|^2$ in Fig. 4(a) and energy dependence of normalized $|I(k'_{||}, k_{||}, \kappa', \kappa, q_z)|^2$ in Fig. 4(b). Figure 4(a) shows that $|\psi_{\mathbf{k}_{||},\kappa}(\mathbf{r})|^2$ decays significantly when $\xi = 0$ where $\kappa$ reaches its maximum, $2m|V_0|/(\hbar^2 g)$. When $|\xi|$ increases, the corresponding $\kappa$ decreases and finally vanishes when $|\xi| = |V_0|$. Remember that $\pm|\xi|$ result in same $\kappa$, therefore the $\xi > 0$ case has a similar trend, which is not shown in this figure. The ITC highly depends on the square of the modules of form factor as shown in Eq. (2). Figure 4(b) shows $|I(k'_{||}, k_{||}, \kappa', \kappa, q_z)|^2$ as a function of the energies of initial state ($\xi$) and final state ($\xi'$) for long phonon wavelength limit, $q_z \to 0$, when $k_z = k'_z = k_F$. We find that $|I(0,0, \kappa', \kappa, 0)|^2$ in the band gap changes slightly when the energy is deep inside the band gap, $|\xi'| \ll |V_0|$ and $|\xi| \ll |V_0|$. Significant increase of the form factor only occurs when $|\xi'| \sim |V_0|$ and $|\xi| \sim |V_0|$. The metals we considered here have similar Fermi energies which are about 1eV below the gap center as shown in Table I. Therefore, the form factor with Fermi energy, $\xi \sim \xi' \sim -1$eV, should dominate the ITC.

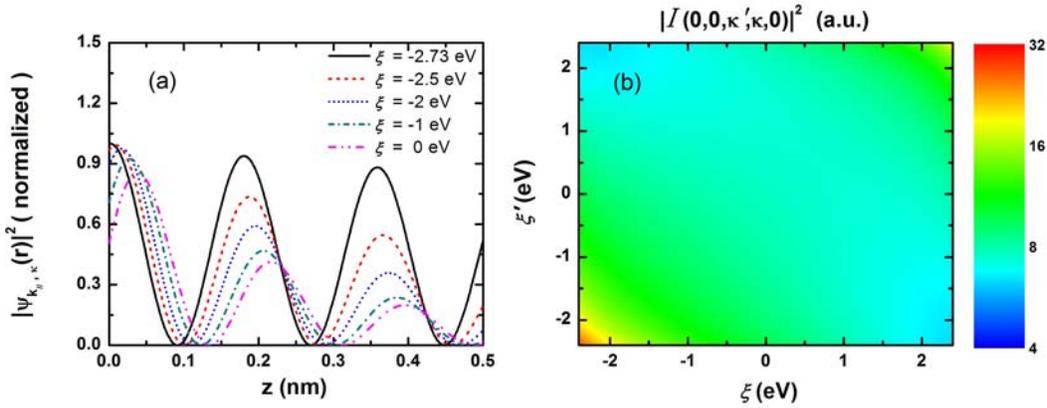

*FIG. 4 (color online) (a) Spatial variations of normalized modular square of the SS for different energy $\xi$. $|\psi_{\mathbf{k}_{||},\kappa}(\mathbf{r})|^2$ decays more significantly when the energy is at the middle of the band gap than that with the energy near the band edge. (b) Modular square of form factor, $|I(0,0, \kappa', \kappa, 0)|^2$, as a function of $\xi$ and $\xi'$. Changes of $|I(0,0, \kappa', \kappa, 0)|^2$ are significant when $|\xi'| \sim |V_0|$ and $|\xi| \sim |V_0|$ while are slight when $|\xi'| \ll |V_0|$ and $|\xi| \ll |V_0|$.*

## IV. SUMMARY

In summary, we have calculated the ITC across metal-insulator/semiconductor interfaces by considering the electron-phonon interaction between surface state electrons and phonons in insulator/semiconductor. The calculated ITC across Pb/Pt/Al/Au-diamond interfaces are very close among these metals though the electronic structures of these metals are quite different. The main reason is the pinning of the Fermi energy in the band gap. This finding explains well the experimental results obtained by Stoner et al.[9] and by Hohensee et al.[11]

## APPENDIX: EXPRESSIONS OF FORM FACTORS

For free phonon modes, the square of the modules of the form factor can be expressed as $|I(k'_{||}, k_{||}, \kappa', \kappa, q_z)|^2 = |B|^2 |B'|^2 \Xi / 16$. In the calculation of Eq. (5), the normalization factor $B$ should be determined by matching the electron wave functions at z=0. For semi-infinite metal, we have $A \to \sqrt{2/L_M}$ and $|B| \to A|\sin\eta / \cos(\phi/2)|$. $|B|$ is determined by $\eta$ and $\phi$ which can be calculated by solving[22]

$$e^{i\phi} = \cos\phi + i\sin\phi = \left(\xi + i\sqrt{V_0^2 - \xi^2}\right)/V_0, \quad \text{(A1a)}$$

$$\tan\eta = \frac{-2k_z}{g\tan\left(\frac{\phi}{2}\right) + 2\kappa}. \quad \text{(A1b)}$$

Here $\phi = 2n\pi + \tan^{-1}\left(\sqrt{V_0^2 - \xi^2}/\xi\right)$ for $\xi \leq 0$ and $\phi = (2n+1)\pi + \tan^{-1}\left(\sqrt{V_0^2 - \xi^2}/\xi\right)$ for $\xi > 0$ where $n$ is arbitrary integer. $|B'|$ and $\phi'$ can be calculated similarly by changing $\xi$ to $\xi'$.

We note $\tilde{\kappa} = \kappa + \kappa'$, $\gamma_1 = q_z^2 + \tilde{\kappa}^2 - g^2$, $\gamma_2 = g\tilde{\kappa}$, $\gamma_3 = q_z^4 + \tilde{\kappa}^4 - g^2\tilde{\kappa}^2 - g^2q_z^2 + 2q_z^2\tilde{\kappa}^2$ and $\gamma_4 = 2g\tilde{\kappa}^3 + 2gq_z^2\tilde{\kappa}$. $\Xi = \Xi_1 + \Xi_2 + \Xi_3 + \Xi_4$ can be expressed in terms of

$$\Xi_1 = \frac{1}{(q_z+g)^2+\widetilde{\kappa}^2} + \frac{1}{(q_z-g)^2+\widetilde{\kappa}^2} + \frac{2}{q_z^2+\widetilde{\kappa}^2}, \tag{A2a}$$

$$\Xi_2 = \frac{2\gamma_1\cos(\phi+\phi')-4\gamma_2\sin(\phi+\phi')}{\gamma_1^2+4\gamma_2^2}, \tag{A2b}$$

$$\Xi_3 = \frac{2\cos(\phi-\phi')}{q_z^2+\widetilde{\kappa}^2}, \tag{A2c}$$

$$\Xi_4 = 4\frac{[\gamma_3(q_z^2+\widetilde{\kappa}^2)+\gamma_2\gamma_4](\cos\phi+\cos\phi')-[\gamma_4(q_z^2+\widetilde{\kappa}^2)-\gamma_2\gamma_3](\sin\phi+\sin\phi')}{\gamma_3^2+\gamma_4^2}. \tag{A2d}$$

For localized phonon modes, $|M(q_{||},q_z)|^2|I(k'_{||},k_{||},\kappa',\kappa,q_z)|^2$ in Eq. (3) can be rewritten into two parts:

$$\frac{|M(q_{||},q_z)|^2}{Q^2}\left\{q_{||}^2\left|\text{Im}[I(k'_{||},k_{||},\kappa',\kappa,q_z)]\right|^2 + q_z^2\left|\text{Re}\left(I(k'_{||},k_{||},\kappa',\kappa,q_z)\right)\right|^2\right\}, \tag{A3}$$

where

$$\left|\text{Im}[I(k'_{||},k_{||},\kappa',\kappa,q_z)]\right|^2 = |B|^2|B'|^2\Xi_{||}/16, \tag{A4a}$$

$$\left|\text{Re}[I(k'_{||},k_{||},\kappa',\kappa,q_z)]\right|^2 = |B|^2|B'|^2\Xi_z/16. \tag{A4b}$$

$\Xi_{||} = q_z^2(\Xi_{||,1}+\Xi_{||,2}+\Xi_{||,3}++\Xi_{||,4})$ can be expressed in terms of

$$\Xi_{||,1} = \frac{2\cos(\phi+\phi')(\gamma_1^2-4\gamma_2^2)-8\gamma_1\gamma_2\sin(\phi+\phi')}{(\gamma_1^2+4\gamma_2^2)^2}, \tag{A5a}$$

$$\Xi_{||,2} = \frac{2[\cos(\phi-\phi')+1]}{(q_z^2+\widetilde{\kappa}^2)^2}, \tag{A5b}$$

$$\Xi_{||,3} = \frac{2}{(\gamma_1^2+4\gamma_2^2)}, \tag{A5c}$$

$$\Xi_{||,4} = \frac{8\cos\left(\frac{\phi-\phi'}{2}\right)\left[\gamma_1\cos\left(\frac{\phi+\phi'}{2}\right)-2\gamma_2\sin\left(\frac{\phi+\phi'}{2}\right)\right]}{(\gamma_1^2+4\gamma_2^2)(q_z^2+\widetilde{\kappa}^2)}, \tag{A5d}$$

and $\Xi_z = \Xi_{z,1}+\Xi_{z,2}+\Xi_{z,3}+\Xi_{z,4}$ can be expressed in terms of

$$\Xi_{z,1} = \frac{2\cos(\phi+\phi')[(\gamma_1^2-4\gamma_2^2)(\widetilde{\kappa}^2-g^2)+8\gamma_1\gamma_2^2]}{(\gamma_1^2+4\gamma_2^2)^2} + \frac{2\sin(\phi+\phi')\left[2\gamma_2(\gamma_1^2-4\gamma_2^2-4\gamma_1\gamma_2(\widetilde{\kappa}^2-g^2))\right]}{(\gamma_1^2+4\gamma_2^2)^2}, \tag{A6a}$$

$$\Xi_{z,2} = \frac{2\widetilde{\kappa}^2[\cos(\phi-\phi')+1]}{(q_z^2+\widetilde{\kappa}^2)^2}, \tag{A6b}$$

$$\Xi_{z,3} = \frac{2(g^2+\widetilde{\kappa}^2)}{(\gamma_1^2+4\gamma_2^2)}, \tag{A6c}$$

$$\Xi_{z,4} = \frac{8\cos\left(\frac{\phi-\phi'}{2}\right)\widetilde{\kappa}\left[(2\gamma_2 g + \gamma_1 \widetilde{\kappa})\cos\left(\frac{\phi+\phi'}{2}\right) + (\gamma_1 g - 2\gamma_2 \widetilde{\kappa})\sin\left(\frac{\phi+\phi'}{2}\right)\right]}{(\gamma_1^2 + 4\gamma_2^2)(q_z^2 + \widetilde{\kappa}^2)}. \tag{A6d}$$

## ACKNOWLEDGEMENTS

JZ would like to thank Prof. D. Cahill for discussions. This work is supported in part by the National Natural Science Foundation of China (Grant No. 11334007), the program for New Century Excellent Talents in Universities (JZ, Grant No. NCET-13-0431), and the Program for Professor of Special Appointment (Eastern Scholar) at Shanghai Institutions of Higher Learning (JZ, Grant No. TP2014012). RY acknowledges the support from DARPA (Grant No. FA8650-15-1-7524).